\documentclass[twocolumn,showpacs,superscriptaddress,preprintnumbers,amsmath,amssymb,prl]{revtex4}
\usepackage[dvips]{graphicx,color}
\usepackage{graphicx}
\usepackage{bm}
\begin{document}
\title{Two-dimensional Heisenberg behavior of $J_\mathrm{eff}=1/2$ isospins in the paramagnetic state of spin-orbital Mott insulator Sr$_2$IrO$_4$}
\author{S.\ Fujiyama}
\email{fujiyama@riken.jp} 
\affiliation{RIKEN Advanced Science Institute, Wako 351-0198, Japan}
\author{H.\ Ohsumi}
\affiliation{RIKEN SPring-8 Center, Sayo, Hyogo 679-5148, Japan}
\author{T.\ Komesu}
\affiliation{RIKEN SPring-8 Center, Sayo, Hyogo 679-5148, Japan}
\author{J.\ Matsuno}
\affiliation{RIKEN Advanced Science Institute, Wako 351-0198, Japan}
\author{B.J.\ Kim}
\affiliation{Department of Advanced Materials, University of Tokyo, Kashiwa 277-8561, Japan}
\altaffiliation{Present address of BJK: Material Science Division, Argonne National Laboratory, Argonne, IL 60439, USA}
\author{M.\ Takata}
\affiliation{RIKEN SPring-8 Center, Sayo, Hyogo 679-5148, Japan}
\author{T.\ Arima}
\affiliation{Department of Advanced Materials, University of Tokyo, Kashiwa 277-8561, Japan}
\affiliation{RIKEN SPring-8 Center, Sayo, Hyogo 679-5148, Japan}
\author{H.\ Takagi}
\affiliation{Department of Physics, University of Tokyo, Hongo 113-0033, Japan}
\affiliation{RIKEN Advanced Science Institute, Wako 351-0198, Japan}
\date{March 6, 2012}
\begin{abstract}
Dynamical correlations of $J_\mathrm{eff}=1/2$ isospins in the paramagnetic state of spin-orbital Mott insulator Sr$_2$IrO$_4$ was revealed by resonant magnetic x-ray diffuse scattering. We found two-dimensional antiferromagnetic fluctuation with a large in-plane correlation length exceeding 100 lattice spacings at even 20 K above the mangnetic ordering temperature. In marked contrast to the naive expectation of strong magnetic anisotropy associated with an enhanced spin-orbit coupling, we discovered isotropic isospin correlation that is well described by the two-dimensional $S=1/2$ \textit{quantum Heisenberg} model. The estimated antiferromagnetic coupling constant as large as $J\sim 0.1$ eV that is comparable to the small Mott gap ($<0.5$ eV) points the weak and marginal Mott character of this spin-orbital entangled system.
\end{abstract} 

\maketitle

In magnetic oxides with 3d transition metal ions, the energy scale of spin-orbit coupling (SOC) is usually smaller than those of crystal field splitting and on-site Coulomb repulsion. SOC can be treated as a perturbation, of which a primal role is to give rise to a magnetic anisotropy. In 5d transition metal oxides, however, SOC is more than one order of magnitude larger than that of 3d due to the pronounced relativistic effect, as large as a half eV, and can modify the electronic structure drastically. A spin-orbital Mott insulator that was recently identified in a layered perovskite Sr$_2$IrO$_4$ is a novel state of correlated electrons~\cite{Kim2008,Kim2009}. Here, a strong SOC, inherent to heavy 5d transition metal Ir$^{4+}$, splits the Ir $t_\mathrm{2g}$ bands into a half-filled $J_\mathrm{eff}=1/2$ bands and completely filled $J_\mathrm{eff}=3/2$ bands, which gives rise to a $J_\mathrm{eff}=1/2$ Mott state induced by a moderate Coulomb repulsion, $U$. The uniqueness of such a spin-orbital Mott insulator is that $J_\mathrm{eff}=1/2$ isospins with the wave function  $\frac{1}{\sqrt{3}}\left(|xy,\pm\rangle \pm|yz,\mp\rangle+i|zx,\pm\rangle \right)$ are accommodated in the outermost 5d orbitals and, therefore, exotic magnetic couplings representing the complex spin-orbital states are anticipated~\cite{Jackeli2009,Wang2011}.  
 
\begin{figure}[htb]
\includegraphics*[width=9cm]{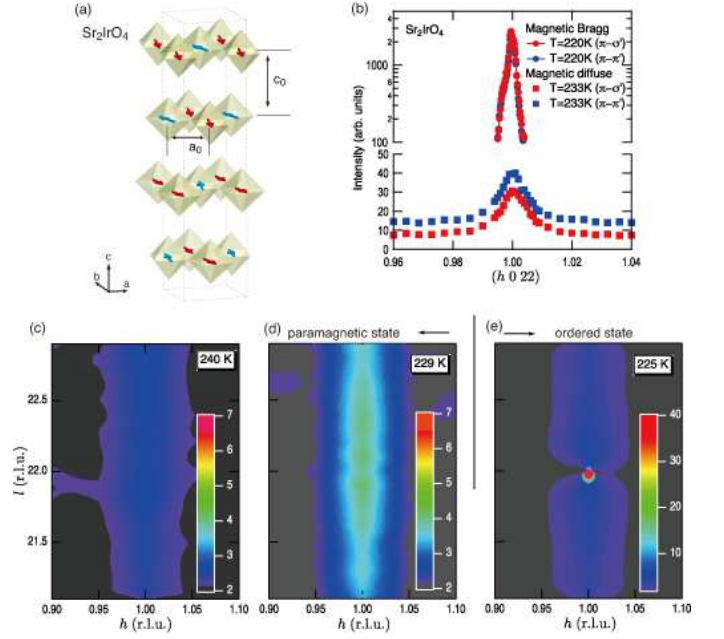}
\caption{(a) Crystal and magnetic structures determined by non-resonant magnetic x-ray scattering of Sr$_{2}$IrO$_{4}$. Magnetic moments lie within the $ab$ plane. (b) Polarization analysis of the Bragg ($T = 220$ K, logarithmic scale) and diffuse ($T = 233$ K) scatterings using $\pi$-$\pi'$ and $\pi$-$\sigma'$ detections by aligning crystalline $\vec{a}+\vec{b}$ direction to the scattering plane. (c)-(e): The intensity maps in the $h$-$l$ plane of the magnetic diffractions from that at 240 K ($T_\mathrm{N} + 11.5$ K) to 225 K ($T_\mathrm{N} - 3.5$ K). Note that the color scale in (e) is shrunk to those in (c) and (d) by a factor of 6, because the intensity of the magnetic Bragg reflection is much larger than that of diffuse scattering. }
\label{fig:fig1}
\end{figure}

Sr$_2$IrO$_4$ is a canted antiferromagnet below 230 K~\cite{Crawford1994,Cao1998,MatsumotoCom}. The $J_\mathrm{eff}=1/2$ moments order approximately antiferromagnetically but significantly canted to cause a ferromagnetic moment of $\sim 0.1\mu_{B}$/Ir within each IrO$_{2}$ ($ab$) plane as shown in Fig.~\ref{fig:fig1}(a). With applying a small magnetic field of $\sim 0.2$ T parallel to the layer, the system shows a metamagnetic transition and becomes a weak ferromagnet. Note that the canting moment is one to two orders of magnitude larger than that of an analogous canted antiferromagnet with 3d copper, La$_2$CuO$_4$. This large canting moment is produced apparently by the interplay between the large SOC and lattice distortion and, at a glance, would imply strongly anisotropy in the isospin coupling.

Recently Jackeli and Khaliullin discussed theoretically the effective Hamiltonian for $J_\mathrm{eff}=1/2$ isospins in Sr$_2$IrO$_4$~\cite{Jackeli2009}. They found that, in the absence of Hund's coupling, the isospin Hamiltonian, including antisymmetric (Dzyloshinskii-Moriya) and symmetric anisotropy terms produced by the interplay between the SOC and the lattice distortion, can be mapped onto the simple Heisenberg Hamiltonian, $\mathcal{H}=\Sigma J S_{i}\cdot S_{j}$ for two-dimensional square lattice by taking proper local axes for each sublattice. Note that, at the strongest limit of SOC, the Land\'e's $g$ factor of the $J_\mathrm{eff}=1/2$ isospin is simply 2 as in the pure $S=1/2$. The conclusion was later supported by Wang and Senthil~\cite{Wang2011}. In reality, however, it is possible that a strong anisotropy, very likely XY of easy plane type, is encountered in the Hamiltonian, for example through a sizable Hund's coupling, and dominates the magnetism. It is therefore not clear at all whether or not Sr$_2$IrO$_4$ with $J_\mathrm{eff}=1/2$ isospins should behave as an $S=1/2$ Heisenberg antiferromagnet. To address this issue experimentally, probing a dynamical correlation among $J_\mathrm{eff}=1/2$ isospins above the ordering temperature should provide us with a definitive clue. However, the conventional approach using neutron scattering is highly constrained due to the strong neutron absorption by Ir ions and, in fact, any clear magnetic peak has not been identified yet in the neutron diffraction pattern.

Magnetic x-ray scattering (diffraction) using synchrotron x-ray source recently emerged as a novel technique to investigate the magnetism of solids. In our previous work, we demonstrated that the resonant magnetic x-ray scattering is particularly powerful to investigate the magnetism of heavy 5d elements such as Ir in the long-range ordered state~\cite{Kim2009}. The $L_\mathrm{II, III}$-edges ($2p\rightarrow 5d$) of 5d elements are located in the hard x-ray region and the short x-ray wavelength at the resonance, $\sim 1 $ \AA, enables us to detect the modulation of spins down to the scale of bond length of Ir's. This is in contrast to the case of 3d magnets, where the $L$-edges are in the soft x-ray region and the wavelength is as long as $\sim 10 $ \AA. In this study, using the marked enhancement at $L_\mathrm{III}$-edge ($2p_{3/2}\rightarrow 5d$), we capture even magnetic diffuse scattering in Sr$_2$IrO$_4$, which represents the instantaneous correlations among $J_\mathrm{eff}=1/2$ isospins.

In this Letter, we disclose the temperature- and $\mathbf{q}$-dependence of spin correlations in a $J_\mathrm{eff}=1/2$ Mott insulator Sr$_2$IrO$_4$ \textit{in the paramagnetic state}, explored by resonant magnetic x-ray \textit{diffuse} scattering. The temperature dependence of in-plane magnetic correlation length well follows that for $S=1/2$ quantum Heisenberg model for two-dimensional square lattice, which implies that the magnetic coupling of the $J_\mathrm{eff}=1/2$ isospins is quite isotropic despite the strong spin-orbit entanglement. The estimated large antiferromagnetic coupling constant as large as $J\sim 0.1$ eV that is comparable to the small Mott gap ($<$ 0.5 eV) points the weak and marginal Mott character through the nonperturbative interplay of spin and charge excitation channels.

A single crystal of Sr$_2$IrO$_4$ with the dimension of 1.2 mm × 0.5 mm × 0.3 mm was grown by a flux method~\cite{Kim2009}. The results of single crystal x-ray diffraction on the crystal were consistent with the space group $I4_{1}/acd$ with tetragonal lattice parameters $a = 5.5$ \AA and $c = 25.8$ \AA~\cite{Cao1998}. Note that $\sqrt{2}\times\sqrt{2}$ superlattice is formed within the plane associated with the staggered rotation of IrO$_{6}$ octahedra and hence $a=\sqrt{2}a_{0}$ ($a_{0} = 3.9$ \AA: the nearest neighbour Ir-Ir distance), as shown in Fig.~\ref{fig:fig1} (a). The unit cell contains four IrO$_{2}$ layers, and we define $c_{0}$ as the distance between the layers with the relation of $c = 4c_{0}$ ($c_{0} = 6.45$ \AA). The magnetization measurements indicated that the antiferromagnetic ordering takes place at a temperature slightly below 230 K in our crystal~\cite{MatsumotoCom}. The magnetic x-ray scattering measurements were performed at SPring-8 BL19LXU, Japan~\cite{Yabashi2001}. The diffuse scattering was detected by a Si-PIN x-ray detector with an energy-resolved multi-channel analyzer. The wavelength of incident x-ray was 1.108 \AA (11.21 keV) corresponding to the Ir $L_\mathrm{III}$-edge ($2p_{3/2}\rightarrow 5d$).

X-ray intensity contour maps of $(h, 0, l)$ plane around (1, 0, 22) at several temperatures are shown in Fig.~\ref{fig:fig1} (c)-(e). At 225 K, a magnetic Bragg reflection (1, 0, 22) is clearly observed as a sharp spot. With increasing temperature to 229 K, the Bragg spot cannot be seen any more but turned into a broad feature associated with diffuse scattering, representing short-range isospin correlations. The change from the Bragg reflection to the diffuse scattering occurred at a temperature between 228 K and 229K, which is consistent with the magnetization measurement. Hereafter we define $T_\mathrm{N} = 228.5 \pm 0.5$ K as the N\'{e}el temperature of the crystal.

The polarization analysis using $\pi$-$\sigma'$ and $\pi$-$\pi'$ detections ~\cite{Gibbs1988} provides a firm evidence of the \textit{magnetic} character of the observed the (1, 0, 22) superspot. The Bragg reflection is described as the square of the scattering length $f_{r}$ as $d\sigma /d\Omega=|f_{r}|^{2}$. Here, $f_{r}$ is a summation of anomalous charge scattering $f_{0}$, magnetic scattering $f_{circ}$, and magnetic linear dichroism $f_{lin}$, that is quadratic in magnetization and negligibly smaller than the former two processes. Of these, $f_{circ}=i(\vec{\epsilon'}\times \vec{\epsilon})\cdot \mathbf{m}\left[F^{1}_{-1}-F^{1}_{+1}\right]$ is relevant to the magnetic scattering originating from electric dipolar transition at the $L_\mathrm{III}$ edge. Here, $\vec{\epsilon}$ $(\vec{\epsilon'})$ shows the polarization vector of the incident (scattered) x-ray, and $F^{1}_{\nu}$ describes the dipolar transition probability which changes the quantum number by $\nu$. The polarization vectors can be rewritten as $\vec{\epsilon}_{\sigma}=\vec{\epsilon}_{\sigma'}=\vec{y}$,  $\vec{\epsilon}_{\pi}=\hat{z}\cos \theta+\hat{x}\sin\theta$ and $\vec{\epsilon}_{\pi'}=\hat{z}\cos\theta-\hat{x}\sin\theta$  by defining a Cartesian $\hat{z}$  as the scattering vector, $\hat{x}$ as $\vec{k}_{i}+\vec{k}_{f}$, $\hat{y}$ as normal to $\hat{x}$ and $\hat{z}$. When we put $\mathbf{m}=m_{x}\hat{x}+m_{y}\hat{y}+m_{z}\hat{z}$ into $f_{circ}$, the expected ratio of the intensities $I_{\pi-\sigma'}/I_{\pi - \pi'}$ is $\left(\sin^{2}\theta \, m_{z}+\cos^{2}\theta \, m_{x}\right)/\sin^{2}2\theta \,m_{y}$. We show in Fig.~\ref{fig:fig1} (b) the Bragg reflections at 220 K using $\pi'$ and $\sigma'$ detections with setting the crystalline $b$-axis as 45$^{\circ}$ shifted from the scattering plane, where $m_{x}=m_{y}=m/\sqrt{2}$ and $m_{z}=0$. The expected ratio of the $\sigma'$ to $\pi'$ channels,  $I_{\pi - \sigma'}/I_{\pi - \pi'}$ , is 1.07, which shows a good agreement with the observed intensities.

The magnetic diffuse scattering above $T_\mathrm{N}$ showed a considerable enhancement only at the $L_\mathrm{III}$ edge as observed for Bragg reflections below $T_\mathrm{N}$~\cite{Kim2009}, which evidences that the $J_\mathrm{eff}=1/2$ state is robust even in the paramagnetic state. Together with the insulating behavior of resistivity above $T_\mathrm{N}$, $J_\mathrm{eff}=1/2$ Mott state is firmly established.

\begin{figure}[htb]
\includegraphics*[width=8cm]{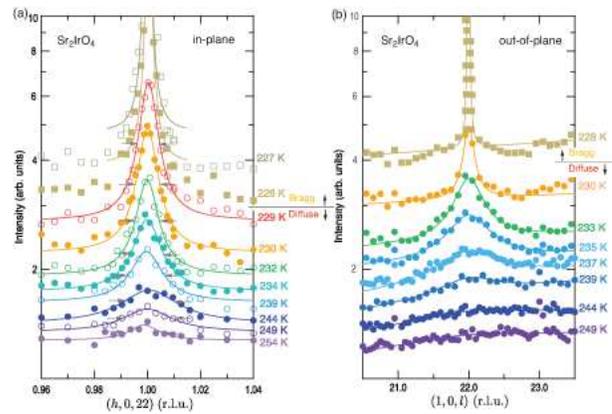}
\caption{Resonant magnetic scattering profiles around (1, 0, 22) magnetic superlattice reflection along the inplane (a) and the out-of-plane (b) directions between 227 K ($= T_\mathrm{N} - 1.5$ K) and 254 K ($= T_\mathrm{N} + 25.5$ K). Baselines of the profiles for $T \le 249$ K (a) and $T \le 244$ K (b) are shifted. Arrows in (a) indicate the half maximum for each diffuse scattering profile.
}
\label{fig:linescan}
\end{figure}

The magnetic diffuse scattering in Fig.~\ref{fig:fig1} (d) is strongly anisotropic, very streaky along the $l$-direction (out-of-plane), apparently indicative of a two-dimensional fluctuation. The in-plane intensity profile $(h, 0, 22)$ and the out-of-plane intensity profile $(1, 0, l)$ are demonstrated in Figs.~\ref{fig:linescan} (a) and (b)~\cite{FitCom}, respectively. Surprisingly, the in-plane profile is sharp and the line-width does not appreciably change with temperature. Even at 244 K, 15.5 K above $T_\mathrm{N}$, the half width at the half maximum of the diffuse peak is as sharp as $\sim 0.01$ r.l.u., which yields a large in-plane correlation length $\xi_a \sim140 a_{0}$ ($a_{0}$: Ir-Ir distance). This implies the presence of a large in-plane coupling, with an energy scale at least substantially larger than the thermal energy $\sim 200$ K. In contrast, at the same temperature of 244 K, the broad diffuse scattering along the \textit{l}-direction with a half width greater than 1 r.l.u. yields an estimate of the out-of-plane correlation length $\xi_{c}<c=4c_{0}$, the magnetic unit-cell length in the ordered state. Those clearly indicate that the correlations are quite two-dimensional and that large two-dimensional antiferromagnetic domains are realized even well above $T_\mathrm{N}$. Upon cooling and critically approaching $T_\mathrm{N} = 228.5$ K, the line profiles along the $l$-direction are rapidly narrowed as shown in Fig.~\ref{fig:linescan} (b), implying contrasting growth of three-dimensional correlations towards the long range ordering. The system, however, appears to remain yet two-dimensional at least down to 1-2 K above $T_\mathrm{N}$, in that critical narrowing of the profile along the in-plane direction is not yet observed.

The observation is in striking parallel to those observed in a layered cuprate La$_2$CuO$_4$, a prototypical $S=1/2$ square-lattice Heisenberg antiferromagnet with a very weak inter-layer coupling. In La$_2$CuO$_4$, the two-dimensional spin correlations dominate above $T_\mathrm{N}$ and the correlation length well develops to over $\sim 100$ lattice spacings, reflecting large in-plane superexchange coupling between neighboring $S=1/2$ spins, $J \sim 0.13$ eV~\cite{Keimer1992}.

\begin{figure}[htb]
\includegraphics*[width=5cm]{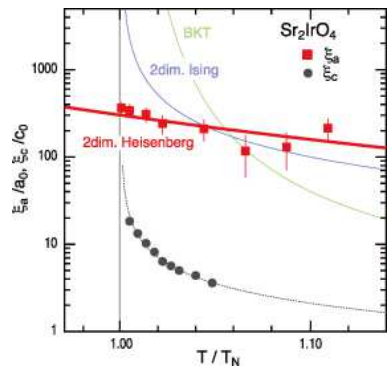}
\caption{Temperature evolutions of in-plane ($\xi_{a}$) and out-of-plane ($\xi_{c}$) spin correlation lengths obtained by a Lorentzian fitting of scattering profiles shown in Fig.~\ref{fig:linescan}. Solid lines are fitted curves to theoretically obtained temperature dependence of $\xi_{a}$ in the square lattice antiferromagnet calculated by $S=1/2$ quantum Heisenberg model (Chakravarty \textit{et al.} and Makivic ~\cite{Chakravarty1989,Makivic1991}) (red), Ising (blue), and XY (green) models.
}
\label{fig:xa}
\end{figure}

To analyze the data further beyond the qualitative argument, we show in Fig.~\ref{fig:xa} the temperature evolution of the in-plane correlation length $\xi_a$ and the out-of-plane $\xi_c$ obtained by the Lorentzian fitting of the peak profiles shown in Figs.~\ref{fig:linescan} (a) and (b)~\cite{Thurston1993}. While the out-of-plane correlation $\xi_c$ appears to show a critical divergence, which can be fitted with $\xi_{c}\propto \left( (T-T_\mathrm{N})/T_\mathrm{N}\right)^{-\nu}$  with $\nu= 0.748$, the in-plane correlation $\xi_a$ shows much more moderate temperature dependence, suggestive of a quantum fluctuation of isospins. If the in-plane magnetic coupling of $J_\mathrm{eff}=1/2$ isospins were to have either strong XY or Ising anisotropy and the long range ordering at $T_\mathrm{N} = 228.5$ K were predominantly driven by such an anisotropy, a Berezinskii-Kosterlitz-Thouless (BKT) transition or a two-dimensional Ising transition would occur. Then, $\xi_a$ would show a critical divergence, following $\xi_{a}\propto \exp \left( 1.5(T-T_\mathrm{N})\right)^{-1/2}$ (BKT) or $\xi_{a}\propto 1/(T-T_\mathrm{N})$ (Ising). The best-fitted curves to these exponential or power-law divergences show fatal contradictions to the observed $\xi_a$, as shown in Fig.~\ref{fig:xa}.

The absence of critical enhancement of $\xi_a$ near $T_\mathrm{N}$ suggests considerable quantum fluctuation of $J_\mathrm{eff}=1/2$ isospins in the paramagnetic state. If the exchange coupling $J$ between the isospins within the $ab$ plane is independent of the directions of the isospins, the observed $\xi_a$ is expected to follow the theoretical formula of the magnetic correlation length, $\xi_\mathrm{2D}(T)$, for $S = 1/2 $ quantum Heisenberg antiferromagnet for two-dimensional square lattice~\cite{Chakravarty1989,Makivic1991}, expressed as
\begin{equation}
\xi_\mathrm{2D}(T)=p\,a_{0}\exp\left(qJ/k_{B}T\right)
\end{equation}

Here, $J$ is the exchange coupling and the constants $p = 0.276$ and $q = 1.25$ are determined by a quantum Monte Carlo study by Makivi\'c for $S=1/2$~\cite{Makivic1991}. The only parameter to fit $\xi_a$ is the antiferromagnetic coupling $J$ between neighboring isospins. As shown in Fig.~\ref{fig:xa}, the fitting works well in the measured temperature range and gives an estimate of  $J = 0.10\pm 0.01$ eV, which is to our surprise almost comparable to that of La$_2$CuO$_4$.

In La$_2$CuO$_4$, the long range ordering is triggered by an inter-layer coupling of the order of $J_\mathrm{perp} \sim 1 \mu$eV, which gives rise to a $T_\mathrm{N} \sim 300$ K for $J \sim 0.13$ eV. If the long range ordering in Sr$_2$IrO$_4$ is understood in terms of 2D $S=1/2$ Heisenberg model, similar parameter sets, $T_\mathrm{N} \sim 228$ K and $J \sim 0.10$ eV, imply an inter-layer coupling of the same magnitude. The analysis of the metamagnetic transition in the ordered state in fact indicates that it should be the case. The energy scale for the interlayer coupling, $J_\mathrm{perp}$, is essentially scaled by the product of the canting moment ($M_{c}$) and the metamagnetic critical field ($H_{c}$), $M_{c}H_{c}$. The experimentally observed value of $M_{c}H_{c}$ is comparable for La$_2$CuO$_4$ ($M_{c} \sim 0.002 \mu_{B}$ and $H_{c} \sim 5$ T) and Sr$_2$IrO$_4$ ($M_{c} \sim 0.06 \mu_{B}$ and $H_{c} \sim0.2 $T) despite contrasting magnitudes of canting moments, evidencing comparable $J_\mathrm{perp}$ in the two systems.

The above analysis indicates that $J_\mathrm{eff}=1/2$ isospins in Sr$_2$IrO$_4$ behave as an $S=1/2$ 2D Heisenberg antiferromagnet despite the strong SOC that is 10 times larger than La$_2$CuO$_4$~\cite{Watanabe2010}, and the magnetic transition is driven by a small interlayer coupling. This unexpected isotropy emerging in the isospins is consistent with the scenario discussed by Jackeli and Khaliullin for the case of zero Hund's coupling~\cite{Jackeli2009}. The effect of Hund's coupling therefore should not be appreciable in Sr$_2$IrO$_4$.

In spite of predominant isotropic character of the isospin correlation, finite anisotropy (XY or Ising-like) is suggested by the polarization analysis of the magnetic diffuse scattering just above $T_\mathrm{N}$ where the observed $\xi_{a}$ shows a small overshoot from the fitted curve. When the magnetic correlation is ideally isotropic and $m_{x}=m_{y}=m_{z}=m/\sqrt{3}$ as a snapshot of fluctuating isospins, the expected ratio of the $\sigma'$ and $\pi'$ detections is $I_{\pi - \sigma'}/I_{\pi - \pi'}=1.40$. The observed diffuse scatterings just above $T_\mathrm{N}$ $(T = T_\mathrm{N} + 4.5$ K), however, have nearly the same intensities as shown in Fig.~\ref{fig:fig1} (b), which is closer to the case of long range ordered state where uniaxial isospin structure is realized. While we can not pin down detail character of the isospin anisotropy, small out-of-magnetization ($m_{z}$) is suggested, which is plausibly originating from a small shift from the ideal $J_\mathrm{eff}=1/2$ manifold by the rotation of IrO$_{6}$ octahedra through exotic spin-lattice couplings.

The large $J \sim 0.1$ eV estimated here poses fundamental questions on the interplay between the ``Mottness'' and the $J_\mathrm{eff}=1/2$ magnetism. In the optical spectra $\sigma (\omega)$ of Sr$_2$IrO$_4$, the absorption peak associated with the excitation across the Mott-Hubbard gap is observed around $E_\mathrm{Mott} \sim 0.5$ eV and the tail of the absorption extends down to 0.2 eV~\cite{Kim2008}, clearly pointing the ``weak'' Mott character. The small optical gap is in fact comparable to the expected height of the dispersion of the spin wave, $2J \sim 0.2$ eV. We have a marginal situation in Sr$_2$IrO$_4$: the exchange coupling, strictly speaking, cannot be treated perturbatively and the coupling for next-nearest-neighbors and further may play a salient role for the magnetism. This is analogous situation to organic quantum spin liquids on triangular lattice~\cite{Kezsmarki2006,Balents2010,Morita2002} but in marked contrast to the case for La$_2$CuO$_4$ where the exchange coupling $J \sim 0.13$ eV is substantially smaller than the Mott-Hubbard (charge transfer) gap $\Delta \sim 2$ eV. The isospin excitation in Sr$_2$IrO$_4$ may meet the continuum of charge excitations at high energies in particular above 100 K where the absorption tail spreads down to well below 0.2 eV, which may alter the nature of high-energy isospin excitations. 

In summary, we demonstrated that the magnetic correlations among $J_\mathrm{eff}=1/2$ isospins in a spin-orbital Mott insulator Sr$_2$IrO$_4$ are described essentially as 2D $S=1/2$ Heisenberg model despite the strong entanglement of spin and orbitals inherent to Ir 5d. A large antiferromagnetic coupling $J \sim 0.1$ eV, comparable even to an $S=1/2$ 3d analogue La$_2$CuO$_4$, was estimated from the temperature dependence of the in-plane correlation length. In the presence of a Mott gap smaller than 0.5 eV, the large $J$ implies a nonperturbative character of exchange interaction associated with a novel coupling between the charge and the isospin excitations. To date, only neutron scattering enables us to study momentum resolved magnetic correlation in solids. We now possess a new technique to investigate spin dynamics by \textit{an electromagnetic wave}. This study together with previous studies~\cite{Kim2009,Ishii2011} establishses that the resonant x-ray technique works particularly well for heavy 5d transition-metal compounds, not only for long-range order but also for short-range correlation, because of the presence of $L_\mathrm{II,III}$-edges in the hard x-ray region.

\begin{acknowledgments}
We are grateful to G. Jackeli, G. Khaliullin, B. Keimer, N. Shannon, J. Kishine, T. Momoi and T. Hanaguri for fruitful discussions. The synchrotron radiation experiments were performed at BL19LXU in SPring-8 with the approval of RIKEN (20080047). This work was supported by Grant-in-Aid for Scientific Research on Priority Areas (19052007, 17071001) from MEXT.
\end{acknowledgments}

\end{document}